\documentclass[preprint,12pt]{elsarticle}

\usepackage{amssymb}
\usepackage{url}
\usepackage{booktabs} 

\usepackage{hyperref}
\usepackage{caption}
\usepackage{rotating}
\usepackage{graphicx}
\usepackage{multirow}
\usepackage{array}
\usepackage{amsmath}

\begin{document}

\begin{frontmatter}




\author[1]{Afzal Ahmed}
\ead{afzal.se@must.edu.pk}
\author[1]{Muhammad Raees}
\ead{raees.se@must.edu.pk}
\affiliation[1]{organization={Mirpur University of Science and Technology},
            city={Mirpur},
            state={AJK},
            country={Pakistan}}

\title{Context-aware Advertisement Modeling and Applications in Rapid Transit Systems}

\begin{abstract}
In today's businesses, marketing has been a central trend for growth. 
Marketing quality is equally important as product quality and relevant metrics. 
Quality of Marketing depends on targeting the right person. 
Technology adaptations have been slow in many fields but have captured some aspects of human life to make an impact.
For instance, in marketing, recent developments have provided a significant shift toward data-driven approaches.
In this paper, we present an advertisement model using behavioral and tracking analysis. 
We extract users' behavioral data upholding their privacy principle and perform data manipulations and pattern mining for effective analysis. 
We present a model using the agent-based modeling (ABM) technique, with the target audience of rapid transit system users to target the right person for advertisement applications.
We also outline the Overview, Design, and Details concept of ABM. 
\end{abstract}


\end{frontmatter}


\section{Introduction}
With the growing usage of technological data-driven advancement in organizational settings, business decision-making needs to be reconsidered. 
Machine learning technologies are becoming prevalent in businesses, however, there are many concerns to make such technologies useful in practical contexts \cite{raees2024explainable}.
This is especially relevant for offline businesses when deciding about optimal strategic and operational marketing and advertising \cite{majid2013context}. 
Due to pervasive mobile and internet usage in modern society, target advertising is an effective strategy to convey the appropriate message instead of randomized spamming, 
Likewise, with anonymous and consensual data collection, companies can advertise to specific market segments meeting the criteria. 
These technologies provide their users with more flexibility concerning when, where, and how to travel.
Understanding the influence of ICTs in our mobile information society \cite{raubal2011cogito, yoon2012social} will be essential for updating environmental policies for our task, and maintaining sustainable mobility and transportation \cite{de2007cell}. 
Moreover, ICTs have provided a wide range of spatio-temporal data sources, which can be used for geographic knowledge discovery and data mining in studies on geographic dynamics, such as human travel behavior and mobility patterns \cite{zhao2015first, chang2016travel}.

Although several spatiotemporal datasets (e.g., geo-referenced mobile phone data) only provide incomplete data with relatively low resolution and few individual attributes, it is important to determine how much and to what extent we can extract knowledge from sparse data sources, as well as dealing with uncertainty in sparse datasets but still the extracted information is very helpful in determining the behavior of a person towards a specific brand.
Due to the widespread use of mobile communications, there have been several studies focusing on acquiring geographical knowledge from geo-referenced mobile phone data and using that knowledge for different purposes. 
A social positioning method (SPM) combines both location data and social attributes of mobile phone users to study the dynamics of urban systems \cite{zhao2015first}.
For instance, by identifying individual trajectories of mobile phone users based on tracked location data, we can capture new insights into understanding the basic law of user trajectories. However social networks are way more expanded these days and almost everyone is connected through this network data extraction from social networks can be quite helpful in determining the behavior of the individual as well as a group of personnel traveling in a specific area and advertising the brands related to them \cite{el2012social}. 
Although the point of consideration while advertising is the customer our proposed method is not only going to help customers but also the companies advertising their brands to only those relevant to them which decreases the cost in terms of money and effort exerted while advertising their brands to those who are not even interested. 

When it comes to shopping many factors affect the behavior of the customer. We have divided the factors into two main states 1) individual state and 2) external/environmental state. The first state is the description of individual activity state focusing on the age group, daily routine, occupation, emotional state income level, etc. while the second state is a description of the external factors affecting the individual’s life in which we shall be focusing on the society, friends circle, and travel patterns, etc \cite{zhao2015first}.
This short article provides a case evaluation of optimizing advertisement strategies to target relevant customers. 
The targeted advertising can be customized to market segments, channels, or geographical areas. 
For instance, by utilizing non-pervasive and anonymous tracking through Information and communication technologies (ICTs), such as mobile phone sensors. 

The remainder of the article is organized in the following pattern section 2 is the literature review conducted to provide the summary of the work already done in fields under discussion section 3 is the proposed methodology to resolve the problem section 4 is the evaluation results to validate our work and section 5 contains the conclusion and future research direction.

\section{Background}
\label{section:2back}
We have reviewed our literature in three main points of view Agent-Based Modeling(ABM), pattern mining and ML applications, and advertisement strategies.

\subsection{Agent based Modeling}
Agent-based modeling is considered to be a powerful simulation modeling technique that helped in solving many problems in the last few years, including applications to real-world business problems \cite{rand2021agent, novosel2015agent, riaz2013lateral}. 
The core objective of using Agent-Based Modeling is to represent the activities and processes involved in a collaborative environment \cite{raees2021context}. 
Agents should be more than just a simple software package or program but it is a bit unclear where the boundaries lie. 
This is the demonstration of a general problem present in AI of defining ‘intelligence’ \cite{raubal2011cogito}.
According to Anumba et al. \cite{anumba2003negotiation} Agents should not only be like that their actions are only the results of their environment but they should have the capability of taking initiative following a goal-directed behavior.
Agent-based modeling is considered an approach that is being used in epidemiological literature \cite{el2012social}. 
Machine learning techniques are quite useful in implementing the behavior of the agents \cite{dehghanpour2016agent}. 
Agent-based models (ABMs), using a complex systems approach, provide a method for examining dynamic interactions of social and demographic actors at both micro and macro levels \cite{williams2017using, raees2021context}.

Haghnevis et al. \cite{haghnevis2016agent} presented a formal agent-based modeling (ABM) platform that enables managers to predict and partially control patterns of behaviors in certain engineered complex adaptive systems (ECASs). 
The approach integrates social networks, social science, complex systems, and diffusion theory into a consumer-based optimization and agent-based modeling (ABM) platform \cite{haghnevis2016agent}.
The Agent-based modeling approaches are used in many fields ranging from economics and social sciences to biology and diverse engineering areas \cite{rand2021agent, dorigo2016swarm, riaz2013lateral}. 
Understanding ABMs important as it should describe a complete model, for instance by standardization. Standardization supports understanding models and making inferences. 
When there are issues in standardization, analyzing the protocol issues is essential and it makes it easier to understand and write models.
A root-cause analysis can also provide an effective strategy to understand what to build in models \cite{raees2020study}.

When the system being modeled is complex, modular decentralized, changeable, and defined at the time of design, ABM is a well-fitted method of modeling \cite{novosel2015agent}. 
Some models about optimal control to determine the speed of transport systems are discussed in the literature \cite{raees2021context}. 
Under the optimal control approach, research studies are mainly grouped into two categories: exact solution algorithm and heuristic algorithm \cite{yang2015survey}. 
Another framework \cite{di2016agent} for cooperation in intermodal freight transport chains as multi-actor systems.
And when it comes to the domain related to the negotiation of the
agents Rosenschein and Zlotkin \cite{rosenschein1994rules} distributed this domain into three subdomains task orientated domains (TODs), state-orientated domains (SODs), and worth-orientated domains (WODs), here each subdomain is a generalization of the previous one TODs are very simple in these the activity of the agents is based on the set of assigned tasks it has to accomplish. 
WODs are the domains where worth is defined by the agents for each
state available. Here goals can be more flexible and making concessions is also allowed. An example would be agents in an e-marketing place where the goal for a company may be to obtain the maximum price for a within period b. 
The chances of deadlocks and conflicts a present here too but here the environment is more bargaining.

\subsection{Pattern Mining}
Majid et al. \cite{majid2013context} stated in their article that the pervasive use of digital cameras and sharing photos on social networks such as Flicker makes a lot of geo-tagged photos
available on the Web. 
Based on the geo-tagged photos and experiences shared over social networks, they got the traveling preference of the user and proposed a new method to recommend user new tourism places relevant to them \cite{majid2013context}.
According to Yoon et al. \cite{yoon2012social} it takes time for the travelers to digest and put together the collected information for use. 
For instance, many Transit systems issue smart cards to traveling
personnel to pay through it. 
Companies mine the data of individuals to reduce the involved complexity in travel patterns and make clusters, thereby, mining customer data for clustering is studied to better service each segment \cite{raees2023four}. 
For instance, businesses can tailor what to market to a specific group of customer based on their existing behavior providing clarity of decision-making for future scenarios. 
Studies use various methods for customer segmentation such as clustering models \cite{raees2023four}.
Bayesian decision tree can also be used to extract the boarding changes in boarding volumes between two transactions conducted consecutively and evaluate this information combined with historical data of the speed profiles extracted using GPS to get the probability of the potential stop \cite{hammawa2011data}. 

Density-based Spatial Clustering of application with noise (DBSCAN) was used to get the travel pattern of individuals efficiently \cite{hammawa2011data,}.
According to Chun et al., \cite{chun2004data} data mining methods based upon the input from past data available for the prediction of the short-term movements of the important currencies, interest rates, or equities have been studied by the researcher to build a quantitative trading tool to improve the trading to make it efficient \cite{chun2004data}. 
Banks have huge databases containing the transactions and other personal details that can be used for predictions but human resources are not enough for this so a data mining technique is necessary to detect the pattern of the customer and predict the profile of the customer \cite{hammawa2011data}.
Liu et al. \cite{liu2003mining} proposed a technique named Mining Data Record (MDR) to mine web page data they extracted the HTML tags to build the tag tree and trained their system on different domains like book, travel, and auction, software and jobs and they also compared the results with other two techniques OMINI and IPEAD and when it came to prediction OMINI only detected 6 out of 46 pages correctly and IPEAD predicted 14 pages correctly while MDR predicted 44 correctly. In predicting patterns IPEAD does not work for similar records but MDR is very good at detecting boundaries \cite{liu2003mining}.

\subsection{Advertisement Strategies}
According to Chen and Tseng \cite{chen2010exploring} the brand equity for a traveler's behavior is dependent on the following three points (1) frequent visits to that place, (2) period spent there during visits, and (3) recommendation frequency of the destination.
In a study Brand equity of a tourist place named Bali was measured in five measurement variables i.e. (1) knowledge about the brand (2) image of the brand (3) associations of the brand (4) observed quality of brand and (5) loyalty to that brand. 
They concluded that maintaining the equity of the brand can help in getting the loyalty of the visitors, choosing the right area to be targeted, and advantage over the competitors \cite{diarta2015influence}. 
Building on Keller’s brand positioning, brand resonance, and brand value chain models, A study has been conducted in which impact
of the brand management process fast-paced advancement in technologies, digital development, and constraints related to environment and social activities have been discussed. 
They concluded their discussion by suggesting that responsiveness according to user needs, innovation in the products, and responsibility are crucial attributes in the management of brand equity. 
They also suggested that global macro changes demand to concentrate on specified brand equity for both the purposes to meet the expectations of the consumers evolving day by day and to be in a competitive state and achieve high performance in the market \cite{gurhan2016customer}.

Krush et al. \cite{krush2013enhancing} explored the connection available between marketing and resources of the sales (e.g. capability of sales and dashboards for marketing) and sensemaking, and the combinational effects on performance of the firm. 
Their study explored that the capability of sales and using dashboards for marketing explicitly contribute to the performance of the firm and it also has an interactive effect with sensemaking. 
In addition to this, sensemaking can affect both growth in the firm and the efficiency of the cost. 
For marketing scholars, sensemaking is important as it plays a vital role in the knowledge capabilities of the firm and makes the firm more successful and sound in facing changes occurring in the market \cite{raees2023four}. 
These studies confirm how important it is for sales and marketing operations to integrate.
Zhao et al. \cite{zhao2015first} discussed the preconditions for the success of the first product launched in the market and the association to existing resources of the firm, and investigated the strategy of the positioning of the product that can mediate the effects of technical resources, marketing resources, and startup expertise of the founding team on the success of the product. 
The authors claimed that the impact of the founding teams with more prior experience is smaller than the founding teams with less prior startup experience.

\section{Context-Aware Advertisement}
\label{section:3method}
\begin{figure}[h]
  \centering
  \includegraphics[height=5cm]{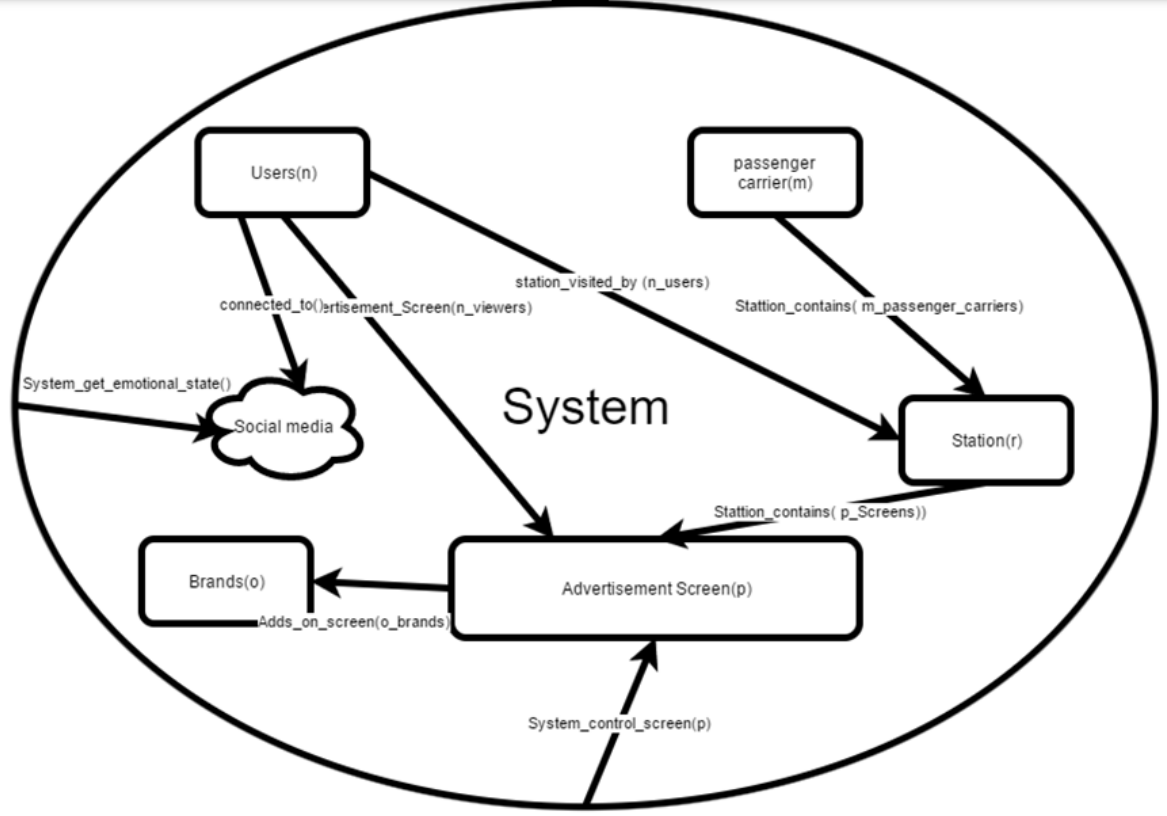}
  \caption{Generic Model of Context-aware Ad. Manager.}
  \label{fig:fig1}
\end{figure}

We used the Agent-Based Modeling (ABM) approach to model our research. In brand management first, we need to identify the agents. 
Our agents are as follows users, Brands, travel vehicles, and stations. A user in our system is connected to the stations he/she travels, the user is also connected to social media where we share our locations our likes and dislikes, and express our emotional states \ref{fig:fig1}. When a person goes to any station he/she will be able to watch ads running on the screen but those ads that are relevant to them are not a big question mark. People ignore such ads which are not relevant to them. We discussed three strategies to improve the ads mechanism to display them on the screen.

\begin{figure}[h]
  \centering
  \includegraphics[height=6cm]{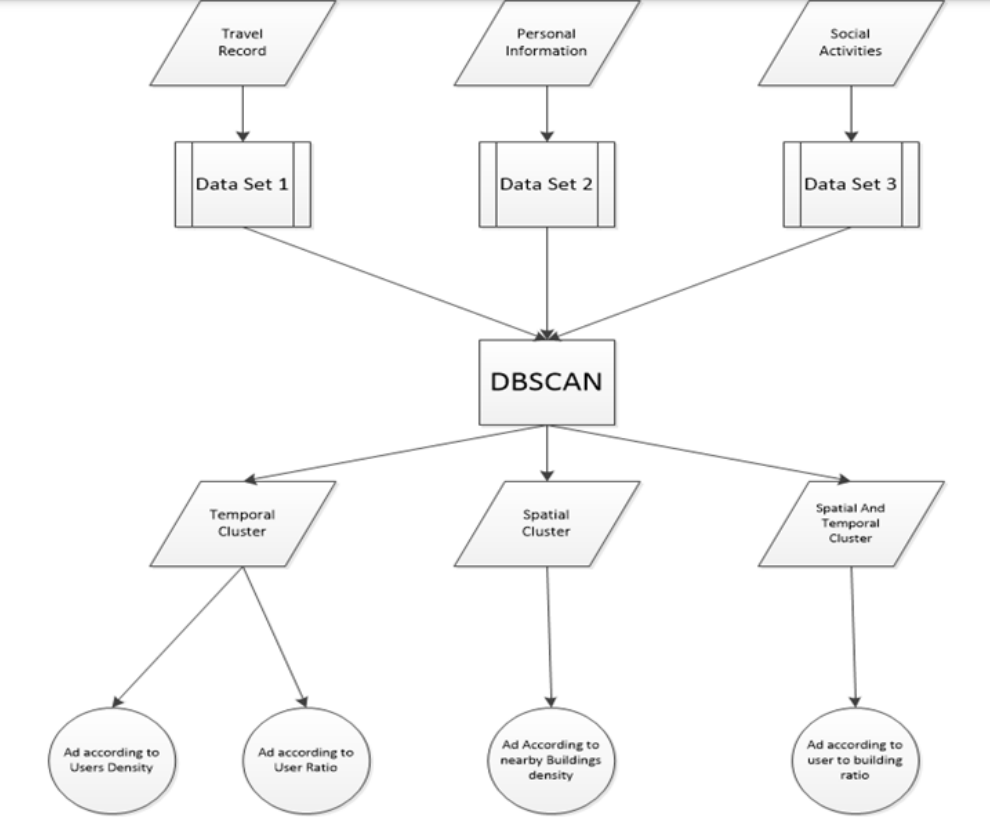}
  \caption{Conceptual Diagram of the Proposed Model.}
  \label{fig:fig2}
\end{figure}

We propose a model \ref{fig:fig2} in which we generate a map against the user's motion getting location data from the GPS of the smartphones they are using and mining that data combined with the social media, station location, brand location, and advertisement screen locations to predict the adds to be displayed on the screen on the different times and different stations.
This model takes user travel history from mobile phone GPS data logs, persons’ personal information from travel company Metro in our case, and social behavior of the person from the links the user has provided while signing up for the Metro card.
All this information is processed by the Density-Based Spatial Clustering of Applications with Noise (DBSCAN). DBSCAN detects the patterns and makes clusters of them. These clusters are then further extracted in two main forms (a) Spatial Clusters: Clusters according to location patterns (b) Temporal Clusters: Clusters according to time patterns followed by the users. Based on these clusters, we divided the ad display tasks into four preferences.

\subsection{Behavioral modeling using ODD model}
\subsubsection{Overview}
\textbf{Purpose:} 
The purpose of the model is to explore the behavior of the people traveling across the stations and to find more effective and efficient ways of the advertisement strategy. Under what conditions the ads shall be displayed on the station screens?
How can the travel patterns combined with their social media activities of people describe the behavior of the people? And how it is helpful in advertisements?

\textbf{Entities, state variables, and scales.} Our model consists of a set of entities, state variables, and scales as described below

\begin{itemize}
    \item Traveling persons Pi
    \item Brands B j
    \item Stations Sk
    \item Each brand and station is characterized by their location Lk
    \item Each traveling personnel has two state variables: his check-in location CIN and check-out location COUT
    \item There is a global variable of hour of the day Th
    \item The distance D between the stations is not specified as we designed a generic model
\end{itemize}

\textbf{Process overview.} The processes in our model are the patterns of the persons traveling from one station to another station at different times, persons visiting the brands, and persons liking the ads. At each time step person moves from one station to another while the station and brands observe at each time step as if some person has visited a specific place.

\subsubsection{Design concept}
In our model, the basic principle that is addressed is the idea of relevant advertisements on the screens.
The screen is going to adapt a specific type of ad at the time according to the changes made in the environment by other agents like the type of persons coming to the station at a specific time. This adaptive behavior is modeled by simply observing the previous travel history of the persons using that track.
Our model does not include direct interaction among the traveling personnel. As at a time traveling toward a station in the peak hours there is a probability that people will be of the same profession. This probability is going to help in the decision-making of the Density-Based Spatial Clustering of Application with Noise (DBSCAN) algorithm and will help make clusters afterward. 

\textbf{Emergence:} This idea emerged from the concept of utilizing rapid transit systems for advertisement. Rapid Transit Systems are one of the most rushy areas during the whole day. Hence, we can target the most appropriate audience at these stations.

\textbf{Observation:} In this model we observe the user movement pattern. The decisions will be made based on user feedback and movement. 

\textbf{Adaptive behavior:} If the user provides positive feedback or remains silent then the ads will continue. Otherwise, more relevant ads will be put to him.

\textbf{Sensing:} We will sense the motion of the users and the frequency of the users. Also, we will sense the trend of the users.

\textbf{Prediction:} Prediction will lead to the decision-making. The system will predict appropriate ad. for specific target.

\textbf{Interaction:} This model will keep the interaction of different stakeholders like a service provider, customer, and broker.

\textbf{Stochasticity:} The movement and task of the users varies. Hence, this system will deal with such variability.

\subsubsection{Model Details}
\textbf{Initialization:} The distance between the stations was initialized at the start of the model. The persons initialized at the start of the model were 1000, stations were initialized to 7, and brands were initialized to 10 at each station. The persons (agents) were initialized according to real-world information about the type of buildings around the stations at the peak hours T1, off-peak hours T2, and night times T3 which have almost 3: 2: 1 ratio respectively.

\textbf{Input:} The brand's feedback will provide the input data to the system.

\textbf{Sub models:} The adapting behavior of the screens sub-model defines what exactly is going to be displayed on the screen it only takes the inputs in the clusters forms and the decision-making is based on the type of times we decided. This sub-model decides whether to display ads according to the density of the people at the station or the density of the buildings surrounding it. The decision of whether to display an ad according to the density of the people at the station or according to the type of the buildings surrounding the station is decided by the variable T h which tells the density of the people at that time of the day at a specific station. Here are some of the preferences we used. 

\textbf{Ads according to Maximum number of audience:} In this technique probability of the number of users will define the type of ads to be displayed on the screen of the station. For example, if there are n total personnel predicted at that time of the day and according to our algorithm let a, b, and c be the number of personnel related to different professions then the ad to be displayed is calculated as

\[
\text{highestO}(a, b, c) \leftrightarrow \text{adToBeDisplayedOnScreen}
\]

\textbf{Ads according to audience ratio:} In this technique ads shall be displayed in the ratio of the audience predicted at that time of the day. For example, if there are n personnel on the station and, a and b are the number of audiences predicted at that time of the day then ads shall be displayed in the ratio of the audiences available. The screen shall be shared by the ads according to the ratio of the audiences detected.

\[
P(a/n) > P(b/n) \leftrightarrow \text{TOS}(a) > \text{TOS}(b)
\]

Here P(a/n) is the probability of a while TOS function returns the Time On Screen value for each ad.

\textbf{Ads According to the nearest Buildings to the station:} In this technique preference is given to the buildings available near the station. A station is mostly filled by the people working in the areas near to that station so it would be a good idea that the ads shall be displayed according to the nearest buildings and the time at which people mostly use the station to go to those buildings. Let d be the set of buildings and t be the set of time spans at which the station is filled with the people going to those buildings.

\[
d(i) \wedge t(j) \leftrightarrow \text{adToBeDisplayedOnScreen}
\]

Where i and j are the indices of the sets a and b and

\[
\text{timeFor}(d(i)) = t(j) \leftrightarrow i = j
\]

\textbf{Ads according to Building ratio:} In this technique ads shall be displayed in the ratio of the building around that station if according to prediction people from the different buildings visit the station at the same time of the day. Then the ads shall be displayed according to the number of people predicatively visiting the station and going to that building. Let d be the set of the building a,b be the number of visitors of that building on the station then mathematically it can be represented as

\[
(a(d(i)) > b(d(j))) \rightarrow ( \text{TOS}(d(i)) > \text{TOS}(d(j)) ) \therefore i \neq j
\]

\textbf{Cost Evaluation:} We evaluate the cost of ads in two main points (1) density of the visitors and (1) density of the buildings near by Cost according to Density of the Visitors. In this case, we calculate the cost of the ads keeping in mind the density of the users in the office starts and end times. Let VR be the relevant user view Let C(VR) be the cost of the relevant view Using these two we can calculate the Total cost of the ad.

\[
\text{TotalCost} = \sum_{i=1}^{n} \text{VR} \cdot C(\text{VR})
\]

Where C(VR) can be calculated as

\[
C(\text{VR}) = (\text{SingleUserPerSecondViewCost}) \times (\text{TotalTimeInSeconds})
\]

\textbf{Cost according to Density of the Nearby Building:} In this case we are going to calculate the cost of the ads being displayed in terms of the building surrounding it as on off days or even office times visitors only visit a station according to the buildings nearby. For this purpose, we need to define the station business factor which is the value of the station at that time of the day. Let, SBR be the Station Business Factor, T be the total time the ad is being displayed, and PR be the per-second rate of the ad then.

\[
\text{TotalCost} = \text{PR} \times \text{SBF} \times \text{CA}
\]

Here CA is the Ad Cost which can be calculated as

\[
CA = \text{PR} \times T
\]

\section{Scenario}
\label{section:4analysis}
Let’s consider the scenario of the Rawalpindi-Islamabad Metro Bus System in Pakistan. This rapid transport system has a total length of 22 kilometers with 24 stations. An average of 0.1 million passengers per day use this transport. Normally, passengers traveling regularly use metro cards. Based on that metro card their daily entering and leaving stations can be recorded.

Suppose, a student of Bahria University Islamabad named Usama has a residence at 6th-Road Satellite Town Rawalpindi. Often, he enters the 6th-Road station between 8:30 am to 9:00 am and checkout at Kechahri station between 8:50 am to 9:20 am. Similarly, Usama enters Kechari station from 5:00 pm to 5:30 pm and leaves at 5:20 pm to 5:50 pm. On weekends and holidays, Usama visits Saddar and 7th Avenue as well.
Depending upon this information we can send him relevant ads. Let’s say Subway has a branch in F-8 and Usama moves in that direction from the nearest station Kechahri. Then, the addition of the Subway will be appropriate for Usama. Similarly, if Usama provides feedback that the Subway advertisement is not relevant to him then, other ads. according to his interest should be sent. Also, Cabinet Computer at 6th Road announces a discount on laptops. So, this ad. should also be sent to Usama. Now, we have a problem that how these ads. will be managed by the broker in case of more appropriate ads. for specific users. For this purpose, our model introduces different schemes. These ads. can be managed based on either the audience ratio or the cost. The formulas described in the model can be used to calculate these factors.

\section{Conclusion}
\label{section:6conclude}
The study has proposed an effective method to improve advertisement strategy. We used travel data of the persons to identify the patterns and to get help in decision-making of which ad to be displayed on any screen. After the evaluation, the results have shown that the travel patterns of the persons are quite helpful in determining the behavior of the person which in turn can be used to make decisions about the advertisement strategy for a specific group of people. We also calculated the cost of the ads to companies according to the number of audiences available at any time of the day.

 \bibliographystyle{elsarticle-num} 
 \bibliography{cas-refs}

\begin{thebibliography}{10}
\expandafter\ifx\csname url\endcsname\relax
  \def\url#1{\texttt{#1}}\fi
\expandafter\ifx\csname urlprefix\endcsname\relax\def\urlprefix{URL }\fi
\expandafter\ifx\csname href\endcsname\relax
  \def\href#1#2{#2} \def\path#1{#1}\fi

\bibitem{raees2024explainable}
M.~Raees, I.~Meijerink, I.~Lykourentzou, V.-J. Khan, K.~Papangelis, From explainable to interactive ai: A literature review on current trends in human-ai interaction, International Journal of Human-Computer Studies (2024) 103301.

\bibitem{majid2013context}
A.~Majid, L.~Chen, G.~Chen, H.~T. Mirza, I.~Hussain, J.~Woodward, A context-aware personalized travel recommendation system based on geotagged social media data mining, International Journal of Geographical Information Science 27~(4) (2013) 662--684.

\bibitem{raubal2011cogito}
M.~Raubal, Cogito ergo mobilis sum: the impact of location-based services on our mobile lives, The SAGE handbook of GIS and society (2011) 159--173.

\bibitem{yoon2012social}
H.~Yoon, Y.~Zheng, X.~Xie, W.~Woo, Social itinerary recommendation from user-generated digital trails, Personal and Ubiquitous Computing 16 (2012) 469--484.

\bibitem{de2007cell}
A.~de~Souza~e Silva, Cell phones and places: The use of mobile technologies in brazil, in: Societies and cities in the age of instant access, Springer, 2007, pp. 295--310.

\bibitem{zhao2015first}
Y.~L. Zhao, D.~Libaers, M.~Song, First product success: a mediated moderating model of resources, founding team startup experience, and product-positioning strategy, Journal of product innovation management 32~(3) (2015) 441--458.

\bibitem{chang2016travel}
Y.~Chang, H.~Zhao-Cheng, Travel pattern recognition using smart card data in public transit, International Journal 6 (2016).

\bibitem{el2012social}
A.~M. El-Sayed, P.~Scarborough, L.~Seemann, S.~Galea, Social network analysis and agent-based modeling in social epidemiology, Epidemiologic Perspectives \& Innovations 9~(1) (2012) 1--9.

\bibitem{rand2021agent}
W.~Rand, C.~Stummer, Agent-based modeling of new product market diffusion: an overview of strengths and criticisms, Annals of Operations Research 305~(1) (2021) 425--447.

\bibitem{novosel2015agent}
T.~Novosel, L.~Perkovi{\'c}, M.~Ban, H.~Keko, T.~Puk{\v{s}}ec, G.~Kraja{\v{c}}i{\'c}, N.~Dui{\'c}, Agent based modelling and energy planning--utilization of matsim for transport energy demand modelling, Energy 92 (2015) 466--475.

\bibitem{riaz2013lateral}
F.~Riaz, S.~I. Shah, M.~Raees, I.~Shafi, A.~Iqbal, Lateral pre-crash sensing and avoidance in emotion enabled cognitive agent based vehicle-2-vehicle communication system, International Journal of Communication Networks and Information Security 5~(2) (2013) 127.

\bibitem{raees2021context}
M.~Raees, T.~A. Khan, K.~Mustafa~Abbasi, A.~Ahmed, S.~Fazilat, I.~Ahmed, Context-aware services using manets for long-distance vehicular systems: A cognitive agent-based model, Scientific Programming 2021~(1) (2021) 8835859.

\bibitem{anumba2003negotiation}
C.~Anumba, Z.~Ren, A.~Thorpe, O.~O. Ugwu, L.~Newnham, Negotiation within a multi-agent system for the collaborative design of light industrial buildings, Advances in Engineering Software 34~(7) (2003) 389--401.

\bibitem{dehghanpour2016agent}
K.~Dehghanpour, M.~H. Nehrir, J.~W. Sheppard, N.~C. Kelly, Agent-based modeling of retail electrical energy markets with demand response, IEEE Transactions on Smart Grid 9~(4) (2016) 3465--3475.

\bibitem{williams2017using}
N.~E. Williams, M.~L. O’Brien, X.~Yao, Using survey data for agent-based modeling: design and challenges in a model of armed conflict and population change, Agent-based modelling in population studies: Concepts, methods, and applications (2017) 159--184.

\bibitem{haghnevis2016agent}
M.~Haghnevis, R.~G. Askin, D.~Armbruster, An agent-based modeling optimization approach for understanding behavior of engineered complex adaptive systems, Socio-Economic Planning Sciences 56 (2016) 67--87.

\bibitem{dorigo2016swarm}
M.~Dorigo, M.~Birattari, X.~Li, M.~L{\'o}pez-Ib{\'a}{\~n}ez, K.~Ohkura, C.~Pinciroli, T.~St{\"u}tzle, Swarm Intelligence: 10th International Conference, ANTS 2016, Brussels, Belgium, September 7-9, 2016, Proceedings, Vol. 9882, Springer, 2016.

\bibitem{raees2020study}
M.~Raees, Study of software quality improvement through reliability metrics models and root cause analysis program, International Journal of Computer Engineering and Information Technology 12~(6) (2020) 42--47.

\bibitem{yang2015survey}
X.~Yang, X.~Li, B.~Ning, T.~Tang, A survey on energy-efficient train operation for urban rail transit, IEEE Transactions on Intelligent Transportation Systems 17~(1) (2015) 2--13.

\bibitem{di2016agent}
A.~Di~Febbraro, N.~Sacco, M.~Saeednia, An agent-based framework for cooperative planning of intermodal freight transport chains, Transportation Research Part C: Emerging Technologies 64 (2016) 72--85.

\bibitem{rosenschein1994rules}
J.~S. Rosenschein, G.~Zlotkin, Rules of encounter: designing conventions for automated negotiation among computers, MIT press, 1994.

\bibitem{raees2023four}
M.~Raees, V.-J. Khan, K.~Papangelis, Four challenges for iml designers: Lessons of an interactive customer segmentation prototype in a global manufacturing company, in: Extended Abstracts of the 2023 CHI Conference on Human Factors in Computing Systems, 2023, pp. 1--6.

\bibitem{hammawa2011data}
M.~Hammawa, Data mining for banking and finance, Oriental J. Comput. Sci. Technol 4 (2011) 273--280.

\bibitem{chun2004data}
S.-H. Chun, S.~H. Kim, Data mining for financial prediction and trading: application to single and multiple markets, Expert Systems with Applications 26~(2) (2004) 131--139.

\bibitem{liu2003mining}
B.~Liu, R.~Grossman, Y.~Zhai, Mining data records in web pages, in: Proceedings of the ninth ACM SIGKDD international conference on Knowledge discovery and data mining, 2003, pp. 601--606.

\bibitem{chen2010exploring}
C.-F. Chen, W.-S. Tseng, Exploring customer-based airline brand equity: evidence from taiwan, Transportation journal 49~(1) (2010) 24--34.

\bibitem{diarta2015influence}
I.~K.~S. Diarta, The influence of bali brand equity on tourists traveling behavior, E-Journal of Tourism 2~(2) (2015) 96--114.

\bibitem{gurhan2016customer}
Z.~G{\"u}rhan-Canli, C.~Hayran, G.~Sarial-Abi, Customer-based brand equity in a technologically fast-paced, connected, and constrained environment, AMS review 6 (2016) 23--32.

\bibitem{krush2013enhancing}
M.~T. Krush, R.~Agnihotri, K.~J. Trainor, E.~L. Nowlin, Enhancing organizational sensemaking: An examination of the interactive effects of sales capabilities and marketing dashboards, Industrial Marketing Management 42~(5) (2013) 824--835.

\end{thebibliography}





\end{document}